\documentclass[%
 amsmath,amssymb,
 aps,
 twocolumn,
reprint,
pra,
nofootinbib
]{revtex4-1}
\usepackage[dvipsnames]{xcolor}

\usepackage[normalem]{ulem}

\usepackage{enumitem}

\usepackage[normalem]{ulem}

\usepackage{graphicx}
\usepackage{dcolumn}
\usepackage{bm}
\usepackage{hyperref}
\hypersetup{colorlinks = true, linkcolor = [rgb]{0.19411,0.51882,0.667058}, urlcolor = [rgb]{0.125490,0.29542,0.1647058}, citecolor = [rgb]{0.75882,0.37411,0.14117}}
\usepackage{makecell}

\usepackage{setspace}
\usepackage{eurosym}
\usepackage{amssymb}
\newtheorem{theorem}{Theorem}
\usepackage{braket}

\graphicspath{{../images/}}
\usepackage[toc,page]{appendix}

\newcommand{\nn}{\nonumber \\}

\usepackage{braket}
\usepackage[bottom]{footmisc}
\def\>{\rangle}
\def\<{\langle}

\usepackage[bottom]{footmisc}
\begin{document}


\title{Fault tolerant quantum data locking}

\author{Zixin Huang} 
\affiliation{Department of Physics and Astronomy, The University of Sheffield, Sheffield, S3 7RH, United Kingdom.}
\author{Pieter Kok} 
\affiliation{Department of Physics and Astronomy, The University of Sheffield, Sheffield, S3 7RH, United Kingdom.}
\author{Cosmo Lupo}
\affiliation{Department of Physics and Astronomy, The University of Sheffield, Sheffield, S3 7RH, United Kingdom.}

\begin{abstract}
Quantum data locking is a quantum communication primitive that allows the use of a short secret key to encrypt a much longer message. It guarantees information-theoretical security against an adversary with limited quantum memory.
Here we present a quantum data locking protocol that employs pseudo-random circuits consisting of Clifford gates only, which are much easier to implement fault tolerantly than universal gates.
We show that information can be encrypted into $n$-qubit code words using order $n - H_\mathrm{min}(\mathsf{X})$ secret bits, where $H_\mathrm{min}(\mathsf{X})$ is the min-entropy of the plain text, and a min-entropy smaller than $n$ accounts for information leakage to the adversary.
As an application, we discuss an efficient method for encrypting the output of a quantum computer.
\end{abstract}

\date{\today}
 
\maketitle

\noindent

\medskip

\section{Introduction}



Quantum data locking (QDL) is a quantum phenomenon that allows us to encrypt a long message using a much shorter secret key with information theoretic security.
This yields one of the strongest violations of classical information theory in quantum physics. In fact, a classic result by Shannon \cite{shannon1949communication}, which is at the root of the one-time pad encryption, establishes that information theoretic encryption of a message of $n$ bits requires a private key of no less than $n$ bits.

The first QDL protocol was introduced by DiVincenzo \textit{et al.}~\cite{DiVin} who showed that a single secret bit is sufficient to obfuscate half of the information contained in $n$ bits, for any $n$.
This was obtained by encoding $n$ bits of classical information into $n$ qubits, where the one bit of information determines which of two mutually unbiased bases is used. Any attempt to measure the $n$-qubit cipher text without knowledge of the basis allows one to obtain at most $n/2$ bits of information.
Further works have strengthened this seminal result \cite{CMP,Aubrun01,Fawzi11,Fawzi13,Dupuis13,Adamczak_2017}. The strongest QDL protocols can encrypt $n$ bits of information using an exponentially small private key, with the guarantee that no more than $\epsilon n$ bits will leak to the adversary.
QDL was discussed in the context of quantum communications in Refs.~\cite{lloyd2013quantum,PRX,Winter2017}, applications to secret key expansion and direct secret communication were introduced in Refs.~\cite{PRL,NJP,QDL_CV}, and proof-of-principle demonstrations were presented in Refs.~\cite{QDL_Pan,PhysRevA.94.022315}.

\begin{figure}[t]
\includegraphics[trim = 0.3cm 0cm 0cm 0cm, clip, width=0.95\linewidth]{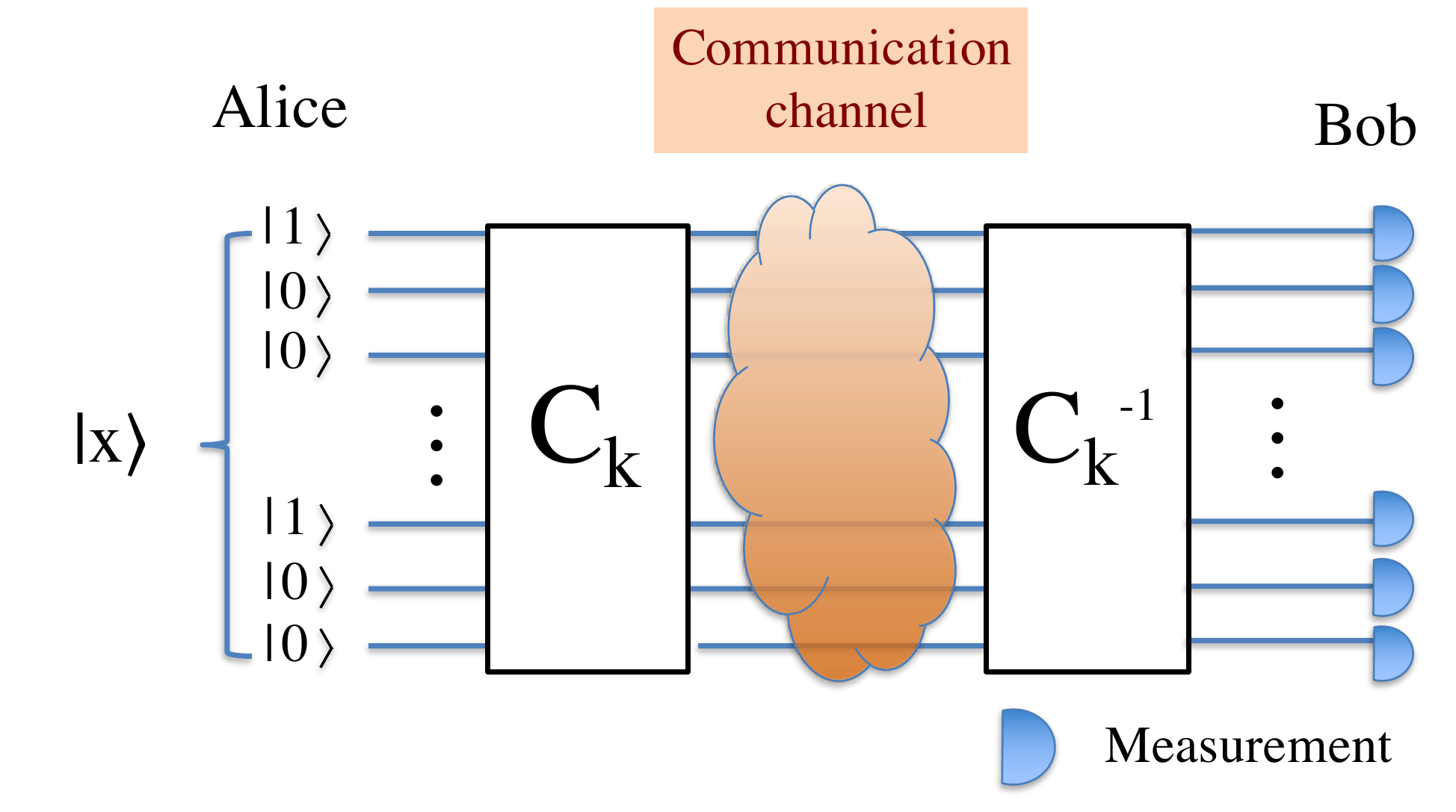} 
 \caption{\label{fig:scheme}Circuit layout for the encryption protocol. 
 A useful state $\ket{x}$ is concatenated with the encryption, a pseudo-random quantum circuit $C_k$. The authorised User applies the unitary $c_k^{-1}$ and correctly decrypts the encryption. An unauthorized User/adversary can attempt to extract information by performing an arbitrary measurement.}
\end{figure}

In a typical QDL protocol, a short private key is used to secretly agree on a code, for example a set of basis vectors, to encode classical information into a quantum system. To encrypt information, the sender (Alice) applies a unitary transformation to map the computational basis into the chosen basis. To decrypt, the legitimate receiver (Bob) applies the inverse transformation, followed by a measurement in the computational basis. This is schematically shown in Fig.~\ref{fig:scheme}.
In order to achieve secure encryption, we require that only a negligible amount of information is obtained by a non-authorized user (Eve) who attempts to measure the quantum cipher text without knowing the private key.
The security of QDL holds independently of the computational capacity of Eve, who may have unlimited computational power, as long as they have limited quantum memory. For example, Eve may have no quantum memory, or a quantum memory with bounded storage time \cite{PRX,PRL,Entropy}. For applications in quantum cryptography, this puts QDL in the framework of bounded quantum storage \cite{QBSM}.

We show that pseudo-random circuits can be used to build QDL protocols that are fault tolerant and robust against information leakage.
In particular, we show that QDL can be realised efficiently using only Clifford gates, which can be made fault tolerant much more easily than the full universal gate set \cite{PhysRevLett.102.110502}. 
We assume that the users have the ability to apply the non-universal set of Clifford gates in a fault-tolerant way. We also assume that the sender can prepare states in the computational basis of $n$ qubits, and the receiver can apply projective measurements in the computational basis.
As an application, we argue that our QDL scheme can be used to encrypt the output of a quantum computer, in such a way that it is accessible only by authorised users.
{This encryption is secure in a scenario where quantum computing is a mature technology but quantum memories are not yet perfect.}


%

The structure of the paper follows. In Sec.~\ref{sec:qdl} we will review the framework of quantum data locking and introduce our protocol. In Sec.~\ref{Sec:random} we discuss the properties of random qubit circuits. We present our security analysis in Sec.~\ref{Sec:definition}, which is followed by our results in Sec.~\ref{Sec:results}. We discuss in detail the application of QDL to securing the output of a quantum computation in Sec.~\ref{Sec:app}.

\section{Quantum data locking}\label{sec:qdl}


Our scheme develops along the lines of previous QDL protocols. 
The protocol involves the legitamate sender Alice and the receiver Bob. The adversary is called Eve.
In QDL, one may distinguish two security scenarios. In \textit{weak} QDL, one assumes that Alice and Bob communicates through a noisy quantum channel, and Eve measures the environment of the channel. This is formally described by saying that Eve has access to the output of the conjugate channel of the channel from Alice to Bob.
In \textit{strong} QDL, one instead assumes that Eve can access the output of both the channel from Alice to Bob and its conjugate.
Here we work in the strong QDL scenario. Furthermore, we consider the case when the channel from Alice to Bob is noiseless. The extension to noisy channels is still an open problem in the general case, with exception of a handful of examples of noisy channels, including the erasure channel and the loss channel \cite{PRL,NJP,QDL_CV,Winter2017}.

The QDL protocol is as follows
\begin{enumerate}

\item Alice and Bob share a unconditionally secure secret key of $\log{K}$ bits.

\item They publicly agree upon a set of $K$ $n$-qubit circuits, $\{ C_k \}_{k=1,\dots,K}$. These circuits are composed of Clifford gates only.

\item Alice encodes the $n$-bit message $x$ into the quantum state $|x\rangle$, which belongs to the $n$-qubit computational basis.

\item She then encrypts the code word and sends it to Bob. The encryption is realised by applying the Clifford circuit corresponding to the unique unitary $C_k$ associated with the private key. Thus, the encrypted code word is $C_k |x\rangle$.

\item Bob, who knows the private key, applies $C_k^{-1}$, decrypts the code word $C_k |x\rangle$, and measures in the computational basis.

\end{enumerate}

Alice can chose one among $M = 2^n$ possible code words. If they have same prior probability, then the code book has maximal entropy of exactly $n$ bits. If the code words do not have equal probabilities, then it is convenient to quantify the entropy of the code book using the min-entropy \cite{Tomamichel}
\begin{align}
    H_\mathrm{min}(\mathsf{X}) = - \log{ p_\mathrm{max} } \, .
\end{align}
where $p_\mathrm{max} := \max_x p_{\mathsf{X}}(x)$. 
A min-entropy smaller than $n$ does also describes a situation where some information about the plain text has leaked to Eve.

The security of QDL is established in a specific setting where the adversary has limited quantum storage capability. For example, Eve may have no reliable quantum memory and thus she is forced to measure the quantum state as soon as she obtains it \cite{PRX}. 
QDL may also be secure if Eve can store quantum information reliably for a limited time, and Alice and Bob have an upper bound on her memory time \cite{PRL,Entropy} \footnote{If Eve has a memory with a finite time, this weakens the security of the protocol, as she may obtain side information during the storage time, and then leverage it to gain more knowledge about the encrypted computation. Past works have addressed this issue in a quantitative way, assuming a model of quantum memory as a noisy channel that decoheres in time \cite{NoisyStorage}. This approach may be used to quantify the security as a function of the time elapsed between when Eve receives the quantum state and when she measures it.}.



A number of QDL protocols and security proofs have been discussed in literature. 
Some of them, however, would be limited to the case where $H_{\mathrm{min}}(\mathsf{X}) = n$ \cite{CMP,Fawzi11,Fawzi13,PRL,NJP}.
For example, Fawzi \emph{et al.}~\cite{Fawzi11,Fawzi13} showed an explicit and efficient construction that can encrypt $n$ bits of information using a key of $O(\log{(n)}\log{(n/\epsilon)})$ bits, with a leakage of no more than $\epsilon n$ bits.
However, this construction cannot be made fault-tolerant~\cite{Fawzi11}.
The approach of Dupuis \emph{et al.}~\cite{Dupuis13} can instead account for non-maximal min-entropy, and would yield results similar to this work, but it relies on sampling unitaries from the Haar distribution, which requires an exponential number of gates~\cite{harrow2018approximate}. In contrast, here we are using an approximate $2$-design, which can be sampled using Clifford gates only.
Finally, the analysis of partial information leakage was also considered in Refs.\ \cite{Robust_phase} as well as in Ref.\ \cite{Dupuis13}, however the scheme of Ref.\ \cite{Robust_phase} may be hard to realize in a fault tolerant way.  

Table~\ref{tab:summary} shows a summary of some previously known QDL protocol, compared to the contribution of this paper, the quantum one-time pad \cite{QOTP}, and the approximate quantum one-time pad \cite{CMP}.

\begin{table}
{
\begin{tabular}{| c || c | c | c |}
\hline	
          					 		& $I_{\rm acc}$  & Key Size & \makecell{Circuit \\ Class} \\ \hline\hline 
\makecell{Quantum\\ one-time \\ pad~\cite{QOTP} }	&  	0								  	 &  $2n$    &  Pauli \\  \hline
\makecell{Approx.\\ quantum \\ one-time \\ pad~\cite{CMP} }    & $\epsilon n$ &  $n + \log n + \log(1/\epsilon^2)$&  Haar \\
\hline
Ref.~\cite{CMP}       & 	$\epsilon n + 3$	  			     & $3 \log n$    &   Haar\\  \hline 
Ref.~\cite{Fawzi13}    &    $ \epsilon n $		 			 	 &	$O(\log(n/ \epsilon)\log n )$  & \makecell{ Universal }   \\  \hline
This paper   &	$ \epsilon n $		  				 & \makecell{$n-H_\text{min}(\mathsf{X}) + $ \\ $ O(\log n ) + O(\log(1/\epsilon))$ }   & Clifford  \\
  \hline  
\end{tabular}}
\caption{\label{tab:summary} Summary of key size and circuit requirement for different schemes for encrypting the information encoded in $n$ qubits.}
\end{table}


\section{Pseudo-random quantum circuits}\label{Sec:random}

Unlike other works, which have considered the uniform ensemble of random unitaries induced by the Haar measure (see e.g.,~\cite{CMP,Fawzi11,Fawzi13}), here we apply pseudo-random unitaries from an approximate 2-design.
This ensemble of unitaries has also been used in other applications related to information obfuscation, most notably system decoupling~\cite{Szehr_2013}.
Using genuine Haar-random unitaries provides slightly more efficient security. However,
as pointed out in Ref.~\cite{brandao2016local}, using unitaries from the Haar measure is prohibitively inefficient for large systems due to the exponential number of two-qubit gates and random bits required.

Recall that, given a Hilbert space of dimensions $d$ and $\delta >0$, a $\delta$-approximate $t$-design is an ensemble of unitary operators $C$ such that~\cite{roy2009unitary,mann2017complexity,brandao2016local}  \begin{align}\label{eq:moments}
(1-\delta) M_\ell \leq \mathbb{E}\left[ \lvert\braket{\alpha|C|\beta}\rvert^{2\ell}\right] \leq 
(1+\delta) M_\ell \, ,
\end{align}
for all unit vectors $\ket{\alpha}$ and $\ket{\beta}$ in $d$ dimensions and $\ell \leq t$,
where $\mathbb{E}$ denotes the expectation value over the $t$-design, and
\begin{align}
M_\ell = \frac{\ell! (d-1)!}{(\ell+d-1)!}
\end{align}
is the $\ell$-th moment of the uniform distribution induced by the Haar measure, i.e.,
$M_\ell = \mathbb{E}_{\mathrm{Haar}}[ \lvert\braket{\alpha|C|\beta}\rvert^{2\ell}]$.

Given an $n$-qubit circuit, a $\delta$-approximate 2-design can be achieved with  $O(n(n+\log{1/\delta}))$ two-qubit Clifford gates \cite{harrow2009random}, or $O(n \log^2n)$ random $U(4)$ gates \cite{harrow2018approximate}. 
%
%
There are known codes that implement the Clifford group in a 
fault-tolerant manner
\cite{zeng2011transversality,PhysRevLett.102.110502}, whereas supplementing the Clifford group with fault tolerant gates into a universal set of gates is highly non-trivial \cite{campbell2017roads}.

The first two moments of the pesudo-random unitaries play an important role in this work, i.e., the first moment  $\mathbb{E}\left[ \lvert\braket{\alpha|C|\beta}\rvert^{2}\right]$,
and the second moment
$ \mathbb{E}\left[\lvert\braket{\alpha|C|\beta}\rvert^{4}\right] \, $.
The ratio
\begin{align} \label{eq:gamma}
\gamma := \frac{\mathbb{E}\left[ \lvert\braket{\alpha|C|\beta}\rvert^{4}\right]} { \mathbb{E}\left[ \lvert\braket{\alpha|C|\beta}\rvert^{2}\right]^2} \, ,
\end{align}
quantifies the spread of the random variable $\lvert\braket{\alpha|C|\beta}\rvert^{2}$ around its average. 
For $\delta$-approximate $2$-designs we can bound $\gamma$ from above as
\begin{align}\label{eq:gammaapp}
\gamma \leq \frac{2 d (1+\delta)}{(d+1) (1-\delta)^2} \leq 2 \frac{1+\delta}{ (1-\delta)^2}\, .
\end{align}
This coefficient will play a fundamental role in our analysis of QDL. We will use the above bound on $\gamma$ to estimate the length of the private key.


\section{Security analysis}\label{Sec:definition}

Our security analysis builds on, improves, and generalises techniques previously applied in Refs.\ \cite{PRL,NJP,huang2019boson}.

Different code words correspond to different quantum states that Alice can prepare, denoted as $\ket{x}$ (with $x=1,...,M$). These vectors are mutually orthogonal. For example, these states can be the vectors in the $n$-qubit computational basis.
Different code words may have different prior probabilities, denoted as $p_{\mathsf{X}}(x)$. 
Therefore, the prior uncertainty in the code words is well quantified by the min-entropy $H_\mathrm{min}(\mathsf{X}) = -\log{\max_x p_{\mathsf{X}}(x)}$.

From the point of view of the legitimate receiver Bob, who knows the private key, the {\it a priori} description of the output of the computation is given by the statistical mixture
\begin{align}
\rho_B = \sum_{x=1}^M p_{\mathsf{X}}(x) \ket{x}\bra{x} \, .
\end{align}
The description of this state is different for Eve, who does not know the private key,
\begin{align} \label{eq:rhoE}
\rho_{E} = \frac{1}{K} \sum_{k=1}^K \sum_{x=1}^M p_{\mathsf{X}}(x)
C_k \ket{x}\bra{x} C_k^\dagger \, .
\end{align} 
Below we show that, if $K$ is large enough, then Eve can obtain  only a negligible amount of information about the code words by measuring $\rho_{E}$.

Like other works on QDL \cite{DiVin,CMP,PRX,PhysRevA.90.022326,PRL,NJP,huang2019boson}, we use the accessible information $I_{\rm acc}(\mathsf{X};E)$ to quantify the potential information leakage to Eve. 
This quantity represents the maximum number of bits of information about the input variable $\mathsf{X}$ that can be obtained from a measurement of the state $\rho_{E}$.
We anticipate that similar results could be obtained using other metrics, see e.g., \cite{Fawzi11,Fawzi13,Adamczak_2017}. 

A measurement is a map $\mathcal{M}_{E \to \mathsf{Y}}$ that takes the quantum system $E$ as input and has the classical variable $\mathsf{Y}$ as output.
For any given measurement, one can consider the mutual information $I(\mathsf{X};\mathsf{Y})$ between the input message and the measurement output. 
Recall that the mutual information between two random variables $\mathsf{X}$ and $\mathsf{Y}$ is $I(\mathsf{X};\mathsf{Y}) = H(\mathsf{Y}) - H(\mathsf{Y}|\mathsf{X})$, where $H(\mathsf{Y}|\mathsf{X})$ is the conditional Shannon entropy. The mutual information vanishes when $\mathsf{X}$ and $\mathsf{Y}$ are statistically independent and reaches its maximum when they are perfectly correlated.
The accessible information is defined as the maximum mutual information,
\begin{align}
I_{\rm acc}(\mathsf{X};E) = \max_{\mathcal{M}_{E \to \mathsf{Y}}} I(\mathsf{X};\mathsf{Y}) \, ,
\end{align}
where the maximization is over all possible measurements $\mathcal{M}_{E \to \mathsf{Y}}$. We require that the accessible information is sufficiently small, i.e., that the information leaking to Eve is negligible not just for one particular measurement, but for all possible measurements she can perform.



The security analysis of the protocol relies on showing that $I_{\rm acc}(\mathsf{X};E)$ can be made arbitrarily small if $K$ is large enough. This also allows us to quantify the minimal length of the private key to ensure secure encryption.
To show this, we first write the accessible information as the difference of two entropies,
\begin{align}\label{2pieces}
I_{\rm acc}(\mathsf{X};E) = \max_{\mathcal{M}_{E \to \mathsf{Y}}} H(\mathsf{Y}) - H(\mathsf{Y}|\mathsf{X}) \, ,
\end{align}
and then show that $H(\mathsf{Y}) \simeq H(\mathsf{Y}|\mathsf{X})$ for all measurements $\mathcal{M}_{E \to \mathsf{Y}}$.
The proof shows that for a random choice of $K$ unitaries, and for $K$ large enough, one obtains $I_{\rm acc}(\mathsf{X};E) \leq 2n\epsilon $ with probability arbitrarily close to $1$.

In general, the measurement map $\mathcal{M}_{E \to \mathsf{Y}}$ is characterised by POVM elements $\Lambda_y$, such that $\Lambda_y \geq 0$, $\sum_y \Lambda_y = \mathbb{I}$. 
It is known that the optimal measurement has unit rank \cite{DiVin}, i.e., the POVM elements take the form $\Lambda_y = \alpha_y |\phi_y\rangle \langle \phi_y|$, where $\phi_y$ are unit vectors, and $\alpha_y$ are positive numbers such that $\sum_y \alpha_y = 2^n$.

The outcomes of the measurement are distributed according to the probability distribution
\begin{align}
p_\mathsf{Y}(y) = \alpha_y \braket{\phi_y|\rho_{E}|\phi_y} \, ,
\end{align}
with $\rho_{E}$ as given in Eq.~(\ref{eq:rhoE}).
For given $x$, the conditional probability of a measurement outcome is 
\begin{align}
p_{\mathsf{Y}|\mathsf{X}=x}(y) = \alpha_y \braket{\phi_y|\rho_{E}^x |\phi_y} \, ,
\end{align}
with 
\begin{align}
\rho_{E}^x = \frac{1}{K} \sum_{k=1}^K C_k |x\rangle\langle x| C_k^\dag \, .
\end{align}

The accessible information in Eq.~(\ref{2pieces}) is then given by
\begin{widetext}
\begin{align}
I_{\rm acc}(\mathsf{X};E) = & \max_{\mathcal{M}_{E \to \mathsf{Y}}} 
\left\{ - \sum_y p_{\mathsf{Y}}(y) \log{p_{\mathsf{Y}}(y)} \right.   \left. + \sum_{xy} p_{\mathsf{X}}(x) p_{\mathsf{Y}|\mathsf{X}=x}(y) \log{p_{\mathsf{Y}|\mathsf{X}=x}(y)} \right\}  \nn
= & \max_{\mathcal{M}_{E \to \mathsf{Y}}} \sum_y \alpha_y \left\{ - \langle \phi_y | \rho_{E} | \phi_y \rangle \log{ \langle \phi_y | \rho_{E} | \phi_y \rangle }  + \sum_x p_{\mathsf{X}}(x) \langle \phi_y | \rho_{E}^x | \phi_y \rangle \log{ \langle \phi_y | \rho_{E}^x | \phi_y \rangle } \right\} \, .
\label{twoterms}
\end{align}

\end{widetext}

The security proof proceeds by showing that, by increasing $K$, both $p_{\mathsf{Y}}$ and $p_{\mathsf{Y}|\mathsf{X}=x}$ concentrate towards their common expectation value, and that the probability of a deviation larger than $\epsilon$ is exponentially suppressed. 
Therefore both the entropy $H(\mathsf{Y})$ and the conditional entropy $H(\mathsf{Y}|\mathsf{X})$ will approach the same value.
We show that both terms in the curly brackets in Eq.~(\ref{twoterms}) are arbitrarily close to 
\begin{align}
\langle \phi_y | \bar{\rho}_{E} | \phi_y \rangle \log{ \langle \phi_y | \bar{\rho}_{E} | \phi_y \rangle }
\end{align}
for all vectors $\phi_y$, where
\begin{align}
\bar\rho_{E} := 2^{-n}    \mathbb{I}
\end{align}
is the $n$-qubit maximally mixed state.
The relative minus sign between the terms then implies that $I(\mathsf{X};\mathsf{Y})$ can be made arbitrarily small. 

First, we show, using the matrix Chernoff bound \cite{Chernoff}, that $\rho_{E}$ is close to the $n$-qubit maximally mixed state $\bar\rho_{E} := 2^{-n} \mathbb{I}$. 
Assuming $K$ is large enough, with near unit probability we have
\begin{align}
\rho_{E} \leq (1+\epsilon) \bar{\rho}_{E} = (1+\epsilon) 2^{-n} \mathbb{I} \, .
\end{align}
From this inequality we find that
$\left \langle \phi \left| \rho_{E} \right | \phi \right \rangle \leq (1+\epsilon) 2^{-n}$ uniformly in $\phi$. 
For a random choice of the unitaries, the probability that this inequality is violated is smaller than (see Appendix \ref{Sec:Chernoff} for details)
\begin{align}\label{p1}
P_1 := \exp{
\left\{
n \ln{2} - K 
\frac{\epsilon^2}{4} \frac{2^{-n}}{p_{\mathrm{\max}}}
\right\}
} \, .
\end{align}

Next, we apply a tail bound due to A. Maurer \cite{AM}. We show that, for given $\phi$ and $x$,
\begin{align}
\langle \phi | \rho_{E}^x | \phi \rangle \geq (1-\epsilon)
\langle \phi | \bar{\rho}_{E} | \phi \rangle \, .
\end{align}
This inequality needs to be extended to all code words and to (almost) all values of $x$. 
In this way we obtain that, for a random choice of the unitaries, the inequality is verified up to a probability smaller than
\begin{align}
P_2 & :=
\exp{\left( 2d\ln{\left(\frac{20 \times 2^n}{\epsilon}\right)} + \frac{\epsilon\ln{M}}{4p_\mathrm{max}} - \frac{ K \epsilon^3 }{128\gamma p_\mathrm{max}} \right)}  \, ,
\label{p2}
\end{align}
where $\gamma$ has been defined in Eq.~(\ref{eq:gamma}) (see Appendix \ref{Sec:Maurer} for details).

Putting these two results together, we obtain 
\begin{align}
I(\mathsf{X};\mathsf{Y}) \leq 2 \epsilon \sum_y \alpha_y 2^{-n} n \, .
\end{align}
Since $\sum_y \alpha_y 2^{-n} = 1$, we finally find
\begin{align}
I(\mathsf{X};\mathsf{Y}) \leq 2 \epsilon n \, .
\end{align}

This bound on the accessible information holds probabilistically, but the likelihood of failure can be made arbitrary small for large enough $K$. 
Specifically, the probability of failure is no larger than $P_1 + P_2$.
Therefore, it can be bounded away from $1$ by chosing $K$ such that
\begin{align}\label{eq:exactk}
K > \max
\begin{cases}
%
\frac{4 n \times 2^n p_{\mathrm{\max}}\ln{2} }{\epsilon^2}  
\, , \cr
\frac{128\gamma}{\epsilon^3}
\left[ 
2^{n+1} p_\mathrm{max} \ln{\left(\frac{20 \times 2^n}{\epsilon}\right)}
+ \frac{\epsilon \ln{M}}{4} 
\right] 
%
\, .
\end{cases}
\end{align}





\section{Results}\label{Sec:results}

\noindent
We have shown that for a random choice of $K$ unitary transformations, the accessible information is upper bounded by a negligible number of bits $2n\epsilon$,
\begin{align}\label{Iacc1}
I_{\rm acc}(\mathsf{X};E) \leq 2n \epsilon \, .
\end{align}
From Eq.~(\ref{eq:exactk}), this holds for a private key of length
\begin{align}\label{K1}
\log{K} = \log{\gamma} + n - H_\mathrm{min}(\mathsf{X}) + 
          O(\log{n}) + O(\log{1/\epsilon}) \, .
\end{align}
Note that the secret key length depends on the coefficient $\gamma$ introduced in Eq.~(\ref{eq:gamma}). For an approximate $2$-design using the bound in Eq.~(\ref{eq:gammaapp}), we obtain
\begin{align} \label{eq:results}
\begin{split}
\log{K} \leq n - H_\mathrm{min}(\mathsf{X}) + \log{\frac{1+\delta}{(1-\delta)^2}} \\
+ O(\log{n}) + O(\log{1/\epsilon}) \, . 
\end{split}
\end{align}
%

\begin{figure}
\includegraphics[trim = 0.3cm 0.5cm 0cm 0cm, clip, width=0.9\linewidth]{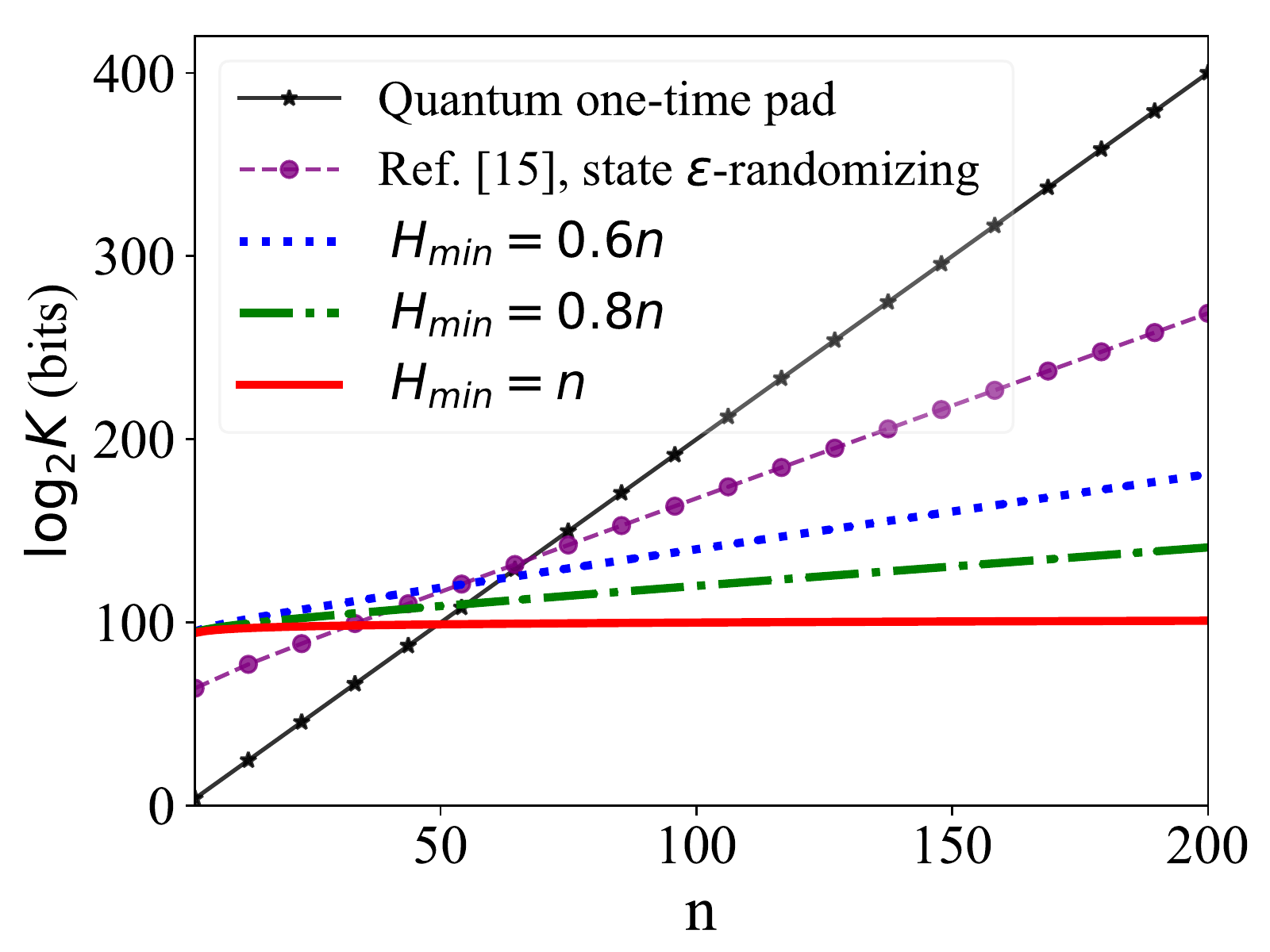} 
 \caption{\label{f:rates} Number of secret bits ($\log K$ in Eq.~\eqref{eq:exactk}) required to lock an $n$-qubit output of a quantum computer, for $\epsilon = 10^{-8}$ and different values of $H_\text{min}$: $H_\text{min}= n$ (red solid line), $0.8 n$ (green dotted-dashed line) and $ 0.6 n$ (blue dotted line). For comparison, we plot the approximate state-randomization in Ref.~\cite{CMP} (purple dashed line with circles), and the quantum one-time pad \cite{CMP} (black line with stars).}
\end{figure}

In conclusions, we have shown that QDL achieves secure encryption using order $n - H_\mathrm{min}(\mathsf{X})$ secret bits, where $H_\mathrm{min}(\mathsf{X})$ is the min-entropy of the code words sent by Alice. 
We plot Eq.~\eqref{eq:results} in Fig.~\eqref{f:rates}, where the exact value of $K$ is given by Eq.~\eqref{eq:exactk}, for $\epsilon=10^{-8}$ 
and different values of $H_\text{min}$. In the figure, we also compare our protocol with other private-key cryptography methods based on the quantum one-time pad as well as its approximate version~\cite{CMP}. %
Our protocol is more efficient, in terms of the length of the private key, when $n \gtrapprox 50$, and the advantage increases with increasing $n$.

\section{Application: securing the output of a quantum computer}\label{Sec:app}

\begin{figure}
\includegraphics[trim = 0cm 0cm 0cm 0cm, clip, width=1.0\linewidth]{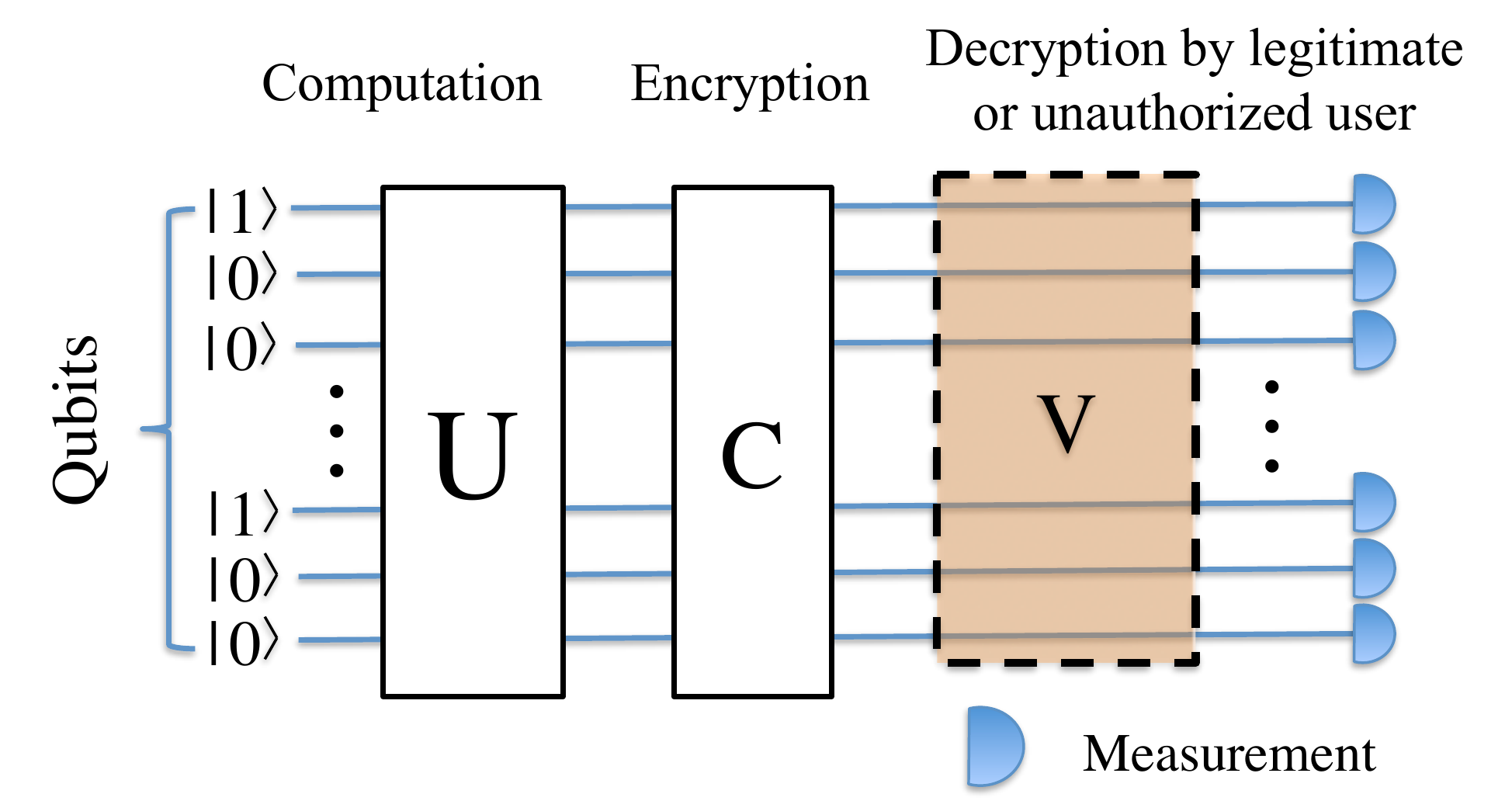} 
 \caption{\label{f:application}  Circuit layout for the encryption protocol. A useful computation $U$ is concatenated with the encryption, a pseudo-random quantum circuit $C$. The authorised User applies the unitary $V=C^\dagger$ and correctly decrypts the encryption. An unauthorized User/adversary can attempt to extract information by performing an arbitrary measurement.}
\end{figure}

Applications of QDL have been mostly focused on quantum communication. Previous works have applied QDL, for example, to communication through a wiretapped channel. 
Here, we propose the use of pseudo-random quantum circuits as efficient encryption devices for protecting the output of a quantum computer.
{This application assumes a scenario where quantum computing is a mature technology but quantum memories are not yet perfect.}

We imagine quantum computers as devices servicing many distributed users, where the latter may have limited computing capability, or may not know the algorithm that is realised by the server. 
In this scenario, we anticipate the need to encrypt the output of a quantum computer.
To realise this task, we consider a protocol for 
private-key encryption between a quantum computer and its authorised user. This is schematically shown in Fig.~\ref{f:application}.
Unlike blind quantum computation \cite{fitzsimons2017private,morimae2012blind,broadbent2009universal,10.1007/978-3-662-45608-8_22}, which is concerned with untrusted hardware and verification, our goal is to prevent unauthorised users from gaining access to the quantum computer's output. 

Otherwise, one could encrypt the quantum state $|\psi\rangle$ before the measurement. Perfect encryption obtained with the quantum one-time pad would require a secret key of $2n$ bits \cite{CMP}.
Approximate encryption, one that encrypts the quantum state up to $\epsilon$ probability of failure, would instead require a secret key of $O(n) + O(\log{1/\epsilon})$ bits \cite{CMP,Aubrun01}.
These protocols require that the encrypted state be virtually indistinguishable from the maximally mixed state. 
Expressed in terms of the trace norm, $\| \rho - 2^{-n}\openone \| \leq \epsilon$, for some small $\epsilon$.
However, the output of a quantum computation typically contains the answer to a meaningful question.
For our purposes, we may simply require that an unauthorised user does not obtain the correct answer.
This opens the possibility of performing the encryption in a much more efficient way.

Suppose the quantum computer is used to solve a particular problem whose solution space has cardinality $M$.
Different outputs of the quantum computation correspond to different quantum states, denoted as $\ket{x}$ (with $x=1,...,M$).
We develop our analysis within the subspace of fault-tolerant computation that incorporates quantum error correction \cite{shor1996fault,lidar2013quantum,roffe2019quantum}. Therefore, the states $\ket{x}$ are assumed to be quantum error-corrected.
For example, during transmission of the quantum state, channel loss will erase a subset of the transmitted qubits. Our protocol allows us to include a redundant encoding to mitigate these losses. As long as the error correction is successful, we know that there is no quantum information leakage, and our protocol remains secure.

We further assume that different outputs are associated with a prior probability $p_{\mathsf{X}}(x)$, and that 
the output states $\ket{x}$ are mutually orthogonal. 
Therefore, the uncertainty in the measurement outcome is well quantified by the min-entropy $H_\mathrm{min}(\mathsf{X}) = -\log{\max_x p_{\mathsf{X}}(x)}$. 
Note that this is the prior distribution of the expected outcome of the computation. 
The value of the min-entropy depends on the particular computation performed by the quantum computer, and it is easy to find examples where $H_{\rm min}$ is low and where it is high. For example, a parity calculation may have a min-entropy as low as $1$, whereas a Grover search may have a min-entropy that is close to maximal.

QDL is particularly efficient when $H_\mathrm{min}(\mathsf{X}) \sim n$, this corresponds to the setting when one has little information about the outcome of the computation.
In this case $n - H_\mathrm{min}(\mathsf{X})$ can be substantially smaller than $n$, suggesting that the encryption can be implemented much more efficiently than previously thought.

\section{Conclusions}\label{Sec:final}

QDL is a communication primitive that allows us to encrypt, with information theoretic security, a long message with a much shorter private key. 
This is impossible in classical information theory, where the key needs to be at least as long as the message. 
When classical information is encoded in a quantum system, the phenomenon of QDL allows for secure encryption against an adversary with limited quantum memory, but unbounded computational power.

In this paper, we have presented a new scheme for QDL that employs pseudo-random unitaries for information scrambling. These unitary transformations belong to an approximate unitary $2$-design. In particular, the unitaries can be obtained by combining Clifford gates. This is an improvement with respect to previous QDL schemes, as fault-tolerant Clifford gates require orders of magnitude fewer physical qubits than universal fault-tolerant quantum computing \cite{campbell2017roads,PhysRevA.95.032338}.
%
Furthermore, our QDL protocol allows for partial information leakage to the eavesdropper. This is modeled by the code words having a non-maximal min-entropy.

We discuss an application of our QDL protocol as a way to encrypt the output of a quantum computer.
Unlike blind quantum computation, which is concerned with untrusted hardware and verification, we focus on preventing unauthorised users gaining access to the output of a quantum algorithm. 
We have considered a scenario where a server can realise fault-tolerant universal quantum computing, the user is capable only of implementing fault-tolerant Clifford gates and measurements in the computational basis, {and the eavesdropper has limited quantum memory.}


%
%

\begin{acknowledgements}\noindent
ZH acknowledges Ryan Mann for fruitful discussions.
We also thank Earl Campell, Armanda Ottaviano-Quintavalle, Joschka Roffe, Omar Fawzi, Dominik Hangleiter,  and the anonymous referees for insightful comments on the manuscript.
This work was supported by the EPSRC Quantum Communications Hub, Grant No. EP/M013472/1.
\end{acknowledgements}

\appendix

\section{Application of the matrix Chernoff bound}\label{Sec:Chernoff}

\noindent
The matrix Chernoff bound states the following (which can be obtained directly from Theorem 19 of Ref.~\cite{Chernoff}):
\begin{theorem}\label{Chernoff}
Let $\{ X_t \}_{t=1,\dots,K}$ be $K$ i.i.d.\ $d$-dimensional Hermitian-matrix-valued random variables, with $X_t \sim X$, 
$0 \leq X \leq R$, and $\mathbb{E}[X] = 2^{-n} \mathbb{I}$.
Then, for any $\epsilon \geq 0$:
\begin{align}
& \mathrm{Pr} \left\{ \frac{1}{K} \sum_{t=1}^K X_t 
\not\leq (1+\epsilon) \mathbb{E}[X] \right\} \nonumber \\
& \leq d \exp{
\left\{
- K D\left[ 
(1+\epsilon)\frac{2^{-n}}{R} \left\| \frac{2^{-n}}{R} \right.
\right]
\right\}
} \, , 
\end{align}
where $\mathrm{Pr}\{ x \}$ denotes the probability that the proposition $x$ is true, and
$D[u\|v] = u \ln{(u/v)} - (1-u) \ln{[(1-u)/(1-v)]}$.
\end{theorem}
Note that for $\epsilon<1$
\begin{align}
    D\left[ 
(1+\epsilon)\frac{2^{-n}}{R} \left\| \frac{2^{-n}}{R} \right.
\right] \geq \frac{\epsilon^2}{4} \frac{2^{-n}}{R} \, .
\end{align}
We apply the Chernoff bound to the $K$ independent random variables 
\begin{align}
X_k \equiv C_k \sum_{x=1}^M p_\mathsf{X}(x) |\psi_x\rangle \langle\psi_x| C_k^\dag \, .
\end{align}
Note that these operators satisfy 
$0\leq X_k \leq p_{\mathrm{max}} := \max_x p_{\mathsf{X}}(x)$. Therefore, $R \equiv p_{\mathrm{\max}}$.
Also note that
\begin{align}
\frac{1}{K} \sum_{k=1}^K X_k
= \frac{1}{K} \sum_{k=1}^K C_k \sum_{x=1}^M p_{\mathsf{X}}(x) |\psi_x\rangle \langle\psi_x| C_k^\dag = \rho_{\mathsf{U}'} \, ,
\end{align}
and $\mathbb{E}[X] = \bar \rho_{\mathsf{U}'} = 2^{-n} \mathbb{I}$.
By applying the Chernoff bound we then obtain
\begin{align}
\mathrm{Pr} \left\{ \rho_{\mathsf{U}'} 
\not\leq (1+\epsilon) 2^{-n} \right\} 
& \leq 2^n \exp{
\left\{
- K 
\frac{\epsilon^2}{4} \frac{2^{-n}}{p_{\mathrm{\max}}}
\right\}
} \\
& = \exp{
\left\{
n \ln{2} - K 
\frac{\epsilon^2}{4} \frac{2^{-n}}{p_{\mathrm{\max}}}
\right\}
} \, .
\end{align}
In conclusion, we have obtained that, up to a probability smaller than 
\begin{align}\label{p1}
P_1 := \exp{
\left\{
n \ln{2} - K 
\frac{\epsilon^2}{4} \frac{2^{-n}}{p_{\mathrm{\max}}}
\right\}
} \, ,
\end{align}
the following matrix inequality holds:
\begin{align}
\rho_{\mathsf{U}'} \leq (1+\epsilon) 2^{-n} \, . 
\end{align}


\section{Application of the Maurer bound}\label{Sec:Maurer}

\noindent
We apply a concentration inequality obtained by Maurer in Ref.~\cite{AM}:
\begin{theorem}\label{Maurer}
Let $\{ X_k \}_{k=1,\dots,K}$ be $K$ i.i.d.\ non-negative real-valued random variables, with $X_k \sim X$ and 
finite first and second moments, $\mathbb{E}[X],\mathbb{E}[X^2] < \infty$.
Then, for any $\tau > 0$ we have that
\begin{align}
\mathrm{Pr}\left\{ \frac{1}{K}\sum_{k=1}^K X_k < (1-\tau) \mathbb{E}[X]  \right\} 
\leq \exp{\left( - \frac{K\tau^2\mathbb{E}[X]^2}{2\mathbb{E}[X^2]} \right)} \, .
\end{align}
\end{theorem}
For any given $x$ and $\phi$, we apply this bound to the random variables
\begin{align}
X_k \equiv | \langle \phi | C_k | \psi_x \rangle|^2 \, .
\end{align}
Note that
\begin{align}
\frac{1}{K} \sum_{k=1}^K X_k = \langle \phi | \rho^x_{\mathsf{U}'} | \phi \rangle \, ,
\end{align}
and 
\begin{align}
\mathbb{E}[X] = \bar \rho_{\mathsf{U}'} = 2^{-n} \mathbb{I} \, .
\end{align}
The application of the Maurer tail bound then yields
\begin{align}\label{pbound1}
\mathrm{Pr}\left\{ \langle \phi | \rho^x_{\mathsf{U}'} | \phi \rangle < (1-\tau) 2^{-n}  \right\} 
\leq \exp{\left( - \frac{K\tau^2 }{2\gamma} \right)} \, .
\end{align}
with $\gamma$ as defined in Eq.~(\ref{eq:gamma}).



The probability bound in Eq.~(\ref{pbound1}) refers to one given value of $x$. 
Here we extend it to $\ell < M$ distinct values $x_1, x_2, \dots, x_\ell$.
We have
\begin{align}\label{ellprob}
& \mathrm{Pr}\left\{  \forall x = x_1, x_2, \dots x_\ell,  \, \, \langle \phi | \rho^{x}_{\mathsf{U}'} | \phi \rangle < (1-\tau) 2^{-n}  \right\} \nonumber \\
& \leq \exp{\left( - \frac{ \ell K\tau^2 }{2\gamma} \right)} \, .
\end{align}
This follows from two observations. First, for different values of $x$, the random variables
$\langle \phi | \rho^{x}_{\mathsf{U}'} | \phi \rangle$ are identically distributed.
Second, these variables are not statistically independent as they obey the sub-normalization constraint 
$\sum_x \langle \phi | \rho^{x}_{\mathsf{U}'} | \phi \rangle = c \leq 1$.
If the $\ell$ random variables $x_1$, $x_2$, $\dots$, $x_\ell$ were statistically independent, then Eq.~(\ref{ellprob}) would hold. However, Eq.~(\ref{ellprob}) still holds because the normalization 
constraint implies that the variables are anti-correlated. 
Therefore, the probability that they are all small is smaller than if they were statistically independent.

We now extend the concentration inequality to all possible choices of $\ell$ values of $x$. 
This amount to a total of ${M \choose \ell}$ events. Applying the union bound we obtain
\begin{widetext}
\begin{align}\label{pbound2}
\mathrm{Pr}\left\{  \exists x_1, x_2, \dots x_\ell,  \, \, | \, \, \forall x = x_1, x_2, \dots x_\ell,  \, \,  \langle \phi | \rho^{x}_{\mathsf{U}'} | \phi \rangle < (1-\tau) 2^{-n}  \right\} 
\leq {M \choose \ell} \exp{\left( - \frac{ \ell K\tau^2 }{2\gamma} \right)} \, .
\end{align}
This implies that up to a probability smaller than ${M \choose \ell} \exp{\left( - \frac{ \ell K\tau^2 }{2\gamma} \right)}$,
$\langle \phi | \rho^{x}_{\mathsf{U}'} | \phi \rangle \geq (1-\tau) 2^{-n}$
for at least $M-\ell$ values of $x$, which yields 
\begin{align} \label{AM10}
\sum_{x=1}^M p_{\mathsf{X}}(x) \langle \phi | \rho^x_{\mathsf{U}'} | \phi \rangle \log{\langle \phi | \rho^x_{\mathsf{U}'} | \phi \rangle} 
\leq \left( \sum_{x \in S_{M-\ell}} p_{\mathsf{X}}(x) \right) (1-\tau) 2^{-n}
\log{(1-\tau) 2^{-n}} \, ,
\end{align}
where $S_{M-\ell}$ denotes the set of $M-\ell$ least likely values of $x$.
Note that 
\begin{align}
\sum_{x \in S_{M-\ell}} p_{\mathsf{X}}(x)
= 1 - \sum_{x \in L_\ell} p_{\mathsf{X}}(x) 
\geq 1 - \ell p_{\mathrm{max}} \, ,
\end{align}
where $L_{\ell}$ is the subset of the $\ell$ most likely values of $x$, and $p_\mathrm{max} = \max_x p_{\mathsf{X}}(x)$.
Putting this into Eq.~(\ref{AM10}) yields
\begin{align}
\sum_{x=1}^M p_{\mathsf{X}}(x) \langle \phi | \rho^x_{\mathsf{U}'} | \phi \rangle \log{\langle \phi | \rho^x_{\mathsf{U}'} | \phi \rangle}
\leq \left( 1 - \ell \, p_\mathrm{max} \right) (1-\tau) 
2^{-n}
\log{(1-\tau) 2^{-n}}
\leq - \left( 1 - \ell \, p_\mathrm{max} \right) (1-\tau) 
2^{-n}
n \, .
\end{align}

Finally, putting $\ell = \tau / p_\mathrm{max}$ we obtain
\begin{align}
\sum_{x=1}^M p_{\mathsf{X}}(x) \langle \phi | \rho^x_{\mathsf{U}'} | \phi \rangle \log{\langle \phi | \rho^x_{\mathsf{U}'} | \phi \rangle}
\leq (1-\tau)^2 2^{-n} n 
= (1-2\tau) 2^{-n} n 
+ O(\tau^2) \, .
\end{align}



To extend to all vectors $\phi$, we exploit the notion of $\delta$-net and closely follows Ref.\ \cite{CMP}.
In this way we obtain
\begin{align}\label{pbound3}
\mathrm{Pr}\left\{ \forall \phi, \exists x_1, x_2, \dots x_\ell,  \, \, | \, \, \forall x = x_1, x_2, \dots x_\ell,  
\, \,  \langle \phi | \rho^{x}_{\mathsf{U}'} | \phi \rangle < (1-2\tau) 2^{-n} \right\} 
\leq \left( \frac{5 \times 2^n}{\tau} \right)^{2d} {M \choose \ell} \exp{\left( - \frac{ \ell K\tau^2 }{2\gamma} \right)} \, .
\end{align}
\end{widetext}

%





%

\end{document}